\documentclass[aps,pra,twocolumn,superscriptaddress,showpacs]{revtex4}

\usepackage{amsmath}
\usepackage{amsfonts}
\usepackage{amssymb}
\usepackage{multirow}
\usepackage{verbatim}
\usepackage{alltt}
\usepackage{moreverb}
\usepackage{graphicx,color,graphics}
\usepackage{hyperref}
\usepackage{setspace}
\usepackage{url}

\providecommand{\ignore}[1]{}

\begin{document}
\singlespacing

\title{Asymptotically optimal data analysis for rejecting local realism}
\author{Yanbao Zhang}
 \affiliation{Department of Physics, University of Colorado at Boulder, Boulder, Colorado, 80309, USA}
 \affiliation{Applied and Computational Mathematics Division, National Institute of Standards and Technology, Boulder, Colorado, 80305, USA}
\author{Scott Glancy}
 \affiliation{Applied and Computational Mathematics Division, National Institute of Standards and Technology, Boulder, Colorado, 80305, USA}
 \author{Emanuel Knill}
 \affiliation{Applied and Computational Mathematics Division, National Institute of Standards and Technology, Boulder, Colorado, 80305, USA}

\date{\today}

\begin{abstract}
Reliable experimental demonstrations of violations of local realism
are highly desirable for fundamental tests of quantum mechanics.  One
can quantify the violation witnessed by an experiment in terms of a
statistical $p$-value, which can be defined as the maximum probability
according to local realism of a violation at least as high as that
witnessed.  Thus, high violation corresponds to small $p$-value.  We
propose a prediction-based-ratio (PBR) analysis protocol whose
$p$-values are valid even if the prepared quantum state varies
arbitrarily and local realistic models can depend on previous
measurement settings and outcomes. It is therefore not subject to the
memory loophole [J. Barrett \emph{et al.}, Phys. Rev. A \textbf{66},
  042111 (2002)]. If the prepared state does not vary in time, the
$p$-values are asymptotically optimal.  For comparison, we consider
protocols derived from the number of standard deviations of violation
of a Bell inequality and from martingale theory [R. Gill,
  arXiv:quant-ph/0110137].  We find that the $p$-values of the former
can be too small and are therefore not statistically valid, while
those derived from the latter are sub-optimal.  PBR $p$-values do not
require a predetermined Bell inequality and can be used to compare
results from different tests of local realism independent of
experimental details.
\end{abstract}


\maketitle

\section{Introduction}
\label{sect:introduction}
Quantum mechanics violates local realism (LR)~\cite{Bell}.  To show
such violation, experimenters usually test a Bell inequality that is
satisfied by all local realistic models (LR models) such as the
Clauser-Horne-Shimony-Holt (CHSH) inequality~\cite{Clauser}
\begin{equation}
\bar{I}_{\text{CHSH}}\equiv E(A_1B_1)+E(A_1B_2)+E(A_2B_1)-E(A_2B_2)\leq2, \label{CHSH}
\end{equation}
where $E(A_iB_j)$ with $i,j \in \{1, 2\}$ is the correlation between
measurements $A_i$ and $B_j$ with outcomes $\pm 1$.  To test this
inequality, each of two parties---Alice and Bob---receives one
particle from a common source. Each performs one of two possible
measurements chosen randomly and independently on their own particle
and records the outcome. We call this procedure a trial. After a large
number of trials, Alice and Bob estimate the CHSH expression
$\bar{I}_{\text{CHSH}}$, which is the left-hand side of the CHSH
inequality, from their joint measurement outcomes. Following this
approach, the departure from LR is typically given in terms of the
number of experimental standard deviations (SDs) separating the
estimate of $\bar{I}_{\text{CHSH}}$ from its LR upper bound of
$2$. For example, Weihs \emph{et al.}~\cite{Weihs} report an
experimental estimate $\tilde{I}_{\text{CHSH}}=2.73\pm0.02$ and claim
a violation of the CHSH inequality by $30$ SDs.

There are several problems with this analysis protocol. First,
although the SD partially characterizes the measurement uncertainty
due to a finite number of trials, it does not consider the probability
that a local realistic system could also violate the inequality after
a finite number of trials.  Because such a system's (non-)violation
can have a larger SD, the experimental SD may suggest a stronger
violation of LR than justified.  Second, one would expect that the
probability distribution of the estimate of $\bar{I}_{\text{CHSH}}$
under LR is Gaussian, since this appears to be justified by the
central limit theorem~\cite{Shao} as the number of trials approaches
infinity. It therefore seems reasonable to statistically quantify the
violation by the probability that a Gaussian random variable can
exceed the mean by the number of SDs of violation experimentally
observed.  However, for a finite number of trials and high violation,
the Gaussianity assumption fails. Third, it is desirable to compare
experimental results from different tests of LR, but the effects of
the problem with experimental SDs and of the failure of Gaussianity
depend on the Bell inequality, the quantum state, measurement
settings, detection efficiency, and other experimental parameters.
Consequently, the number of SDs of violation cannot be used to
directly compare the amount of evidence for rejecting local realism
obtained from different experimental tests.

In this paper, we show how to analyze data from experimental tests of
LR to compute a measure of the strength of the evidence against LR. By
computing this measure, LR violation by different experiments can be
rigorously assessed and compared.  Specifically, the proposed analysis
protocol quantifies LR violation in terms of $p$-values, where small
$p$-values imply strong violation.  We call this the prediction-based-ratio
 (PBR) protocol. Protocols such as this compute a $p$-value from
a ``test statistic'', that is, a value $T(\mathbf{x})$ computed from
the data $\mathbf{x}$.  There are many such statistics to choose from;
an example is the average Bell-inequality violation and is used by the
SD-based protocol. The $p$-value returned by the protocol is computed
from a putative upper bound $b(t)$ on the tail probabilities
$\text{Prob}(T(\mathbf{x})\geq t)$ for $\mathbf{x}$ distributed
according to LR models.  The $p$-value of the protocol given the
observed data $\mathbf{x}$ is defined by $p^{(\text{prot})} =
b(T(\mathbf{x}))$.  In order to be able to interpret the protocol's
$p$-value as a measure of LR violation, it must satisfy statistical
validity: The protocol and its $p$-values are valid if the bound
$b(t)\geq \text{Prob}(T(\mathbf{x})\geq t)$ is true whenever
$\mathbf{x}$ is distributed according to an LR model. See
App.~\ref{sect:statistics_terms} for a discussion of the relevant
statistical concepts and justification for the use of $p$-values.

We prove that the PBR protocol is valid and compare it to SD- and
martingale-based~\cite{Gill1,Gill2} protocols. For $n$ independent and
identically distributed trials, these protocols have the
property that the $p$-value $p$ is exponentially close to $0$.  That
is, $p\simeq 2^{-Gn}$ for large $n$. We call $G$ the asymptotic
confidence-gain rate. It is desirable to have a high confidence-gain
rate as this implies that fewer trials are needed to achieve the same
strength of violation of LR.  The optimal confidence-gain rate that
can be achieved by any protocol is given by the statistical strength
$S$ in units of bits per trial as defined in Ref.~\cite{VanDam}. We
prove that the PBR protocol is asymptotically optimal. That is, its
$p$-values achieve the optimal confidence-gain rate.  The
confidence-gain rates for different protocols are shown in
Figs.~\ref{fig:gain_unbalanced} and~\ref{fig:gain_balanced} for a
number of experimental configurations that are explained in the next
section.  The figures show that SD-based $p$-values are not valid in
some regions. Because the relationship of the SD-based confidence-gain
rates compared to the asymptotically optimal ones varies
substantially, results of experiments with different configurations
cannot be directly compared by the common ``number of SDs of
violation'' measure.  The martingale-based protocol is valid and
computationally simple but has suboptimal confidence-gain rates.

The PBR protocol remains valid if the prepared quantum state varies
arbitrarily and the LR models to be rejected depend on previous
measurement settings and outcomes, that is, in the presence of the
memory effect~\cite{Barrett2}. This is desirable not only for tests of
LR but also for practical applications of quantum information, such as
device-independent quantum key distribution~\cite{Acin2, Masanes},
randomness expansion~\cite{Monroe}, state estimation~\cite{Bardyn} and 
certification of entangled measurements~\cite{Rabelo}.

Compared with the other two protocols, an advantage of the PBR
protocol is that it can be applied to a wide variety of configurations
(the combinations of quantum state, measurement settings and other 
relevant parameters) without having to specify a Bell inequality.  
Since such Bell inequalities characterize the family of setting and 
outcome distributions achievable by LR models, they provide a useful 
guide to designing an experiment and determining good goal configurations 
to be achieved. But since Bell-inequality violation is not directly related
to statistical strength, it is not obvious how to choose the best
inequality before the experiment. Moreover, the predetermined Bell
inequality restricts a successful experiment to configurations close
to the goal, closer than may be achievable in a given experiment.  The
PBR protocol automatically adapts to deviations from the goal,
achieving optimal confidence-gain rates for the actual configuration.
One can exploit this adaptability by applying the PBR protocol to
experiments in progress. This makes it possible to monitor the current
(non-)violation of LR for the purpose of optimizing experimental
parameters and settings.  The online ancillary files 
contain the code and documentation for an
implementation of the PBR protocol (the local realism analysis engine)
that can be used for monitoring experiments in progress and for
analyzing existing data sets. Our results show that the PBR protocol
is sufficiently efficient for practical use with typical experimental
configurations.

The paper is structured as follows: In Sec.~\ref{sect:results}, we
summarize the mentioned methods for calculating $p$-values and show
how their confidence-gain rates compare for tests of LR based on Bell
inequalities. The methods are applied to and compared on simulated and
actual experiments.  The theory for the methods is in
Sec.~\ref{sect:theory}. We assume that the readers are familiar with
the basics of LR and tests of LR based on Bell inequalities. For
reviews of the field, see
Refs.~\cite{Peres,Werner,Genovese,Horodecki}.

\section{Comparison of Protocols}
\label{sect:results}

We consider three protocols that determine $p$-values for LR rejection
from experimental data: SD-based, martingale-based, and PBR protocols.
The first two depend on a Bell inequality, whereas the PBR protocol
requires only the sequence of measurement settings and outcomes.

\begin{figure}[tb!]
   \includegraphics[scale=0.56, viewport=3.3cm 8.9cm 17.3cm 20.8cm]{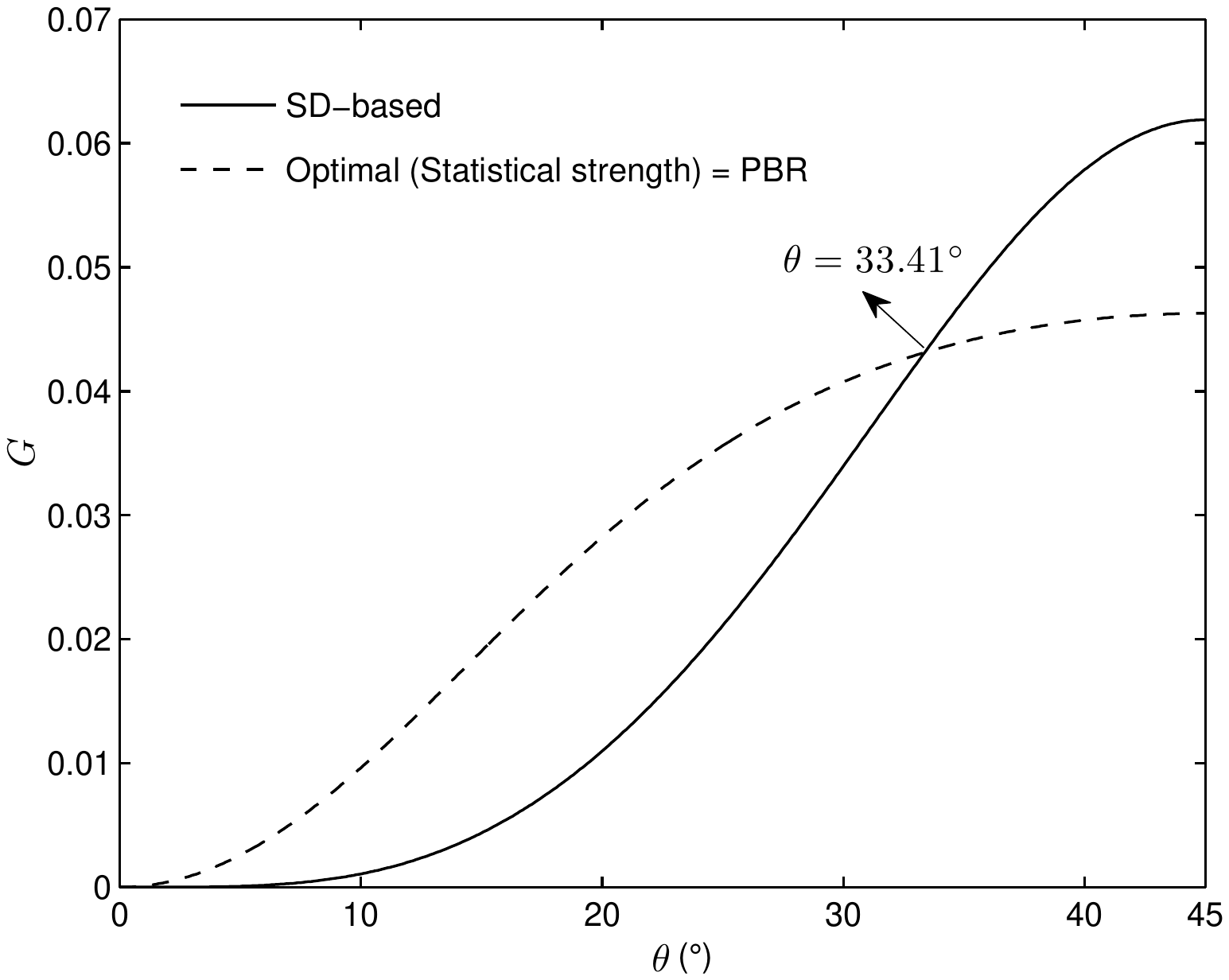}
   \small{
   \caption{Confidence-gain rates $G$ for the SD-based protocol.  $G$
     is shown for a CHSH test of LR with an unbalanced Bell state with
     no loss and perfect detectors. It depends on the parameter
     $\theta$ in the unbalanced Bell state
     $\cos(\theta)|00\rangle+\sin(\theta)|11\rangle$.  The measurement
     settings are chosen to maximize the violation of the CHSH
     inequality~\eqref{CHSH}.  $G$ is compared with the optimal gain
     rate given by the statistical strength
     (Sec.~\ref{subsect:pbr_theory}) for this test. The cross-over
     occurs at $\theta=33.41^\circ$.  SD-based confidence-gain rates
     were computed with respect to the conventional method for
     estimating violation, see Sec.~\ref{subsect:sd_theory}.}
   \label{fig:gain_unbalanced}}
\end{figure}

\begin{figure}[tb!]
   \includegraphics[scale=0.56, viewport=3cm 10cm 24cm 20.7cm]{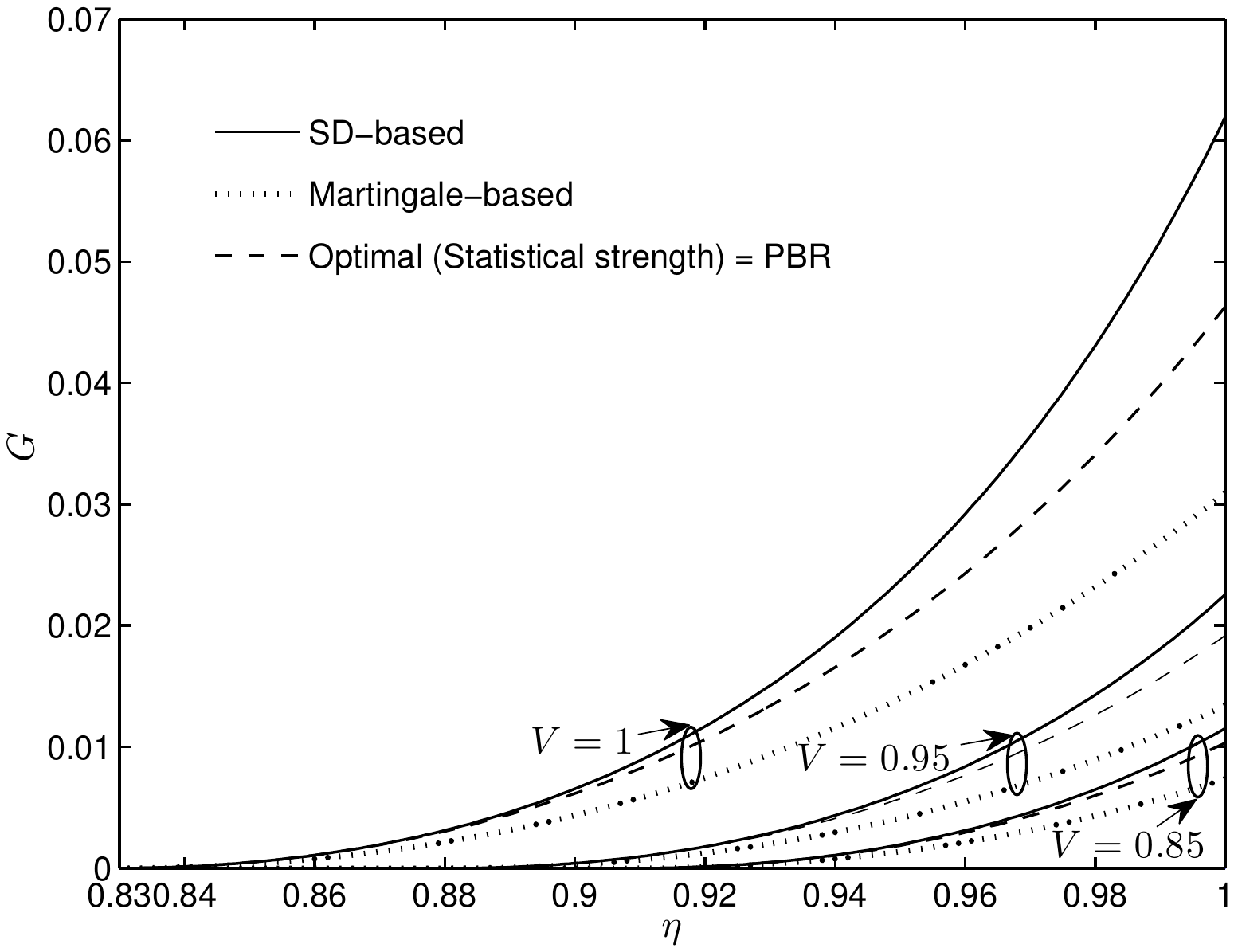}
   \small{
   \caption{The confidence-gain rate $G$ of a CHSH test of LR with
     Bell states and varying detection efficiency $\eta$ and visibility
     \textit{V}.  The measurement settings are chosen to maximize the
     violation of the CHSH inequality~\eqref{CHSH}. Measurement
     outcomes where no photon is detected are assigned the value
     $-1$.}
   \label{fig:gain_balanced}}
\end{figure}

\begin{figure*}[tb!]
   \includegraphics[scale=0.75, viewport=0.5cm 9.3cm 21.5cm 19cm]{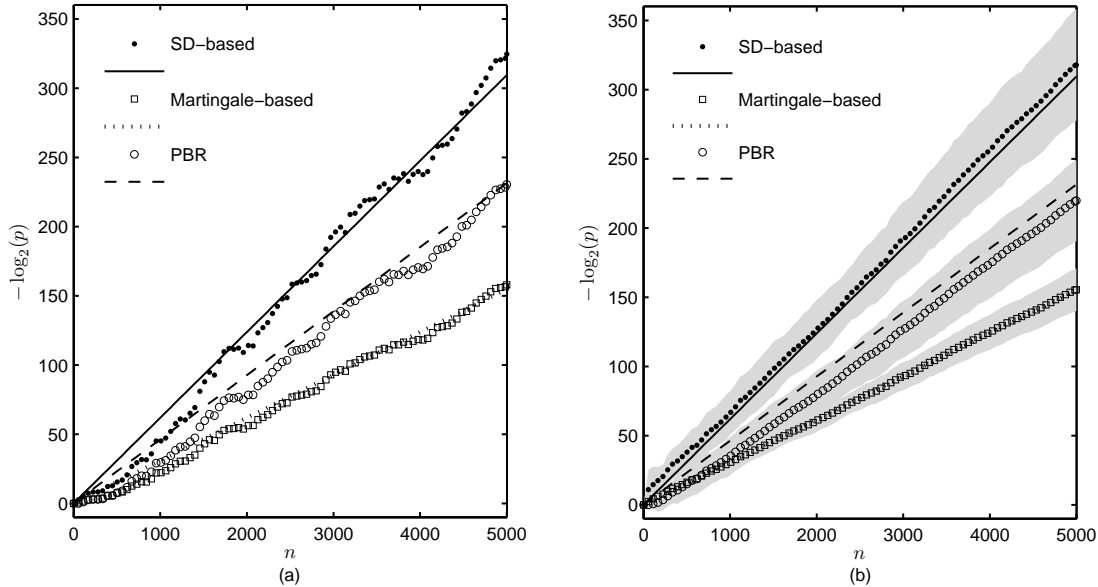}  
   \small{
   \caption{Running log-$p$-values as a function of the number of
     trials $n$ in a CHSH test of LR with a Bell state
     without noise or inefficiency.  The log-$p$-values are computed
     according to the three protocols discussed. The slopes of the straight lines
     are the asymptotic confidence-gain rate for each protocol.
     (a) is for one simulation of $5000$ successive trials. (b) is an
     average of $30$ simulations. The square roots of the unbiased
     estimates of the one-run variances are shown as gray regions
     around the averages and indicate the expected fluctuation for one
     sequence of $n$ trials for each $n$ plotted. Note that for one
     sequence, the fluctuations are not independent as the sequence
     progresses.}
   \label{fig:simulation}}
\end{figure*}

For the purposes of discussion, we fix a Bell inequality
\begin{equation}
\langle I(x)\rangle\leq B,
\label{eq:proper_Bell}
\end{equation}
where $I(x)$ is a real-valued function of the measurement setting and
outcome combination $x$ of a single trial, and $\bar I=\langle
I(x)\rangle$ is its expectation.  Here, the measurement setting
distribution is built into the inequality.  An example is the CHSH
inequality in Eq.~\eqref{CHSH}.  In this case, if $x$'s settings are
$i,j$ and its outcomes are $a,b$, then
\begin{equation}
I(x)=(1-2\delta_{i,2}\delta_{j,2})ab/p_{i,j},\text{ and } B=2,
\label{eq:CHSH_good}
\end{equation}
where $p_{i,j}$ is the probability of choosing the setting combination
$i,j$ in each trial.  The functional form $I(x)$ in
Eq.~\eqref{eq:CHSH_good} ensures that its expectation is equal to the
left-hand side of the CHSH inequality~\eqref{CHSH}. In particular,
this requires dividing by the known probabilities of the measurement
settings. There is no loss of generality by fixing the setting
probabilities in advance. Violation of LR requires that measurement settings be
chosen independently of any hidden variables. In particular, the
locality and memory loopholes cannot be closed unless in each trial,
measurement settings are chosen randomly and independently by each
party with no possibility of a causal connection and according to a
known probability distribution. We allow for arbitrary setting
distributions in Eq.~\eqref{eq:CHSH_good}. For the results in 
Figs.~\ref{fig:gain_unbalanced}, \ref{fig:gain_balanced}, 
\ref{fig:simulation} and~\ref{fig:monroe}, $p_{i,j}=1/4$.

Given an experimentally obtained sequence of settings
and outcomes $x_1,\ldots,x_n$ from $n$ trials, we get an estimate
$\hat I=\frac{1}{n}\sum_{k=1}^n I(x_k)$ of $\bar I$. Note that this
approach differs from the one where each expectation in
Eq.~\eqref{CHSH} is separately estimated by conditioning on the
respective measurement settings, as is commonly done in experiments to
produce an estimate $\tilde I$ of $\bar I$. The difference is
discussed in Sec.~\ref{subsect:sd_theory} and does not significantly
affect the comparisons made here. In this section we outline and
compare the protocols.  Technical details are in
Sec.~\ref{sect:theory}.

\subsection{SD-based Protocol}

The results from the trials are used to obtain $\tilde I$ 
and estimate the SD $\sigma$ of $\tilde I$.  Given that
$\tilde I > B$, it is conventional to give $(\tilde I - B)/\sigma$,
the number of SDs of violation, as a measure of the amount of
violation.  If we pretend that the probability distribution of the
estimate of $\bar I$ given LR is Gaussian with mean bounded by $B$ and
variance $\sigma^2$, we can compute a $p$-value
\begin{equation}
p^{({\text{SD}})}=Q\left(\frac{\tilde I-B}{\sigma}\right),
\label{eq:confidence_sd}
\end{equation}
where $Q(z)$ is the $Q$-function, which is the probability that a 
standard normal random variable $N$ satisfies $N\geq z$.  As a 
function of the number of trials $n$, $\sigma\sqrt{n}$ approaches 
$\sigma_1$, where $\sigma_1$ is an effective one-trial SD.
For large $n$, the quantity $Q((\tilde I-B)/\sigma)$ approaches
$e^{-n(\bar I - B)^2/(2\sigma_1^2)}$.  Thus the asymptotic
confidence-gain rate for the SD-based protocol is
\begin{equation}
G_{\text{SD}} = \log_2(e)\frac{(\bar I - B)^2}{2 \sigma_1^2}.
\end{equation}

SD-based $p$-values are not valid because the experimental SD is
different from the worst-case SD assuming LR, and because deviations
from Gaussianity in the extreme tail of the distribution for $\tilde
I$ cannot be asymptotically neglected.  To explain this issue, define
the random variable $F=\sqrt{n}(\tilde I-B)/\sigma_1$.  For any LR
model, $\langle F\rangle \leq 0$.  We expect that according to the
central limit theorem, $F-\langle F\rangle$ converges in distribution
to a standard normal distribution.  Assuming LR models have the same
or a smaller SD, we are interested in the probability of the event
that $F\geq \sqrt{n} V_n/\sigma_1$, where $V_n$ is the violation of
the Bell inequality found after $n$ trials. But convergence in
distribution cannot be used to compute probabilities of events that
depend on $n$.

A comparison of the confidence-gain rate for the SD-based protocol to
the asymptotically optimal one is shown in
Fig.~\ref{fig:gain_unbalanced}. It implies that SD-based $p$-values
can be lower than justified and are therefore not valid. The worst
case is when the state used is a Bell state, i.e., a maximally 
entangled state of two qubits, which is an aim of most
experiments to date.  The family of unbalanced Bell states considered
in Fig.~\ref{fig:gain_unbalanced} is of interest because they are more
tolerant of low detection efficiency~\cite{Eberhard}.  Experimental
techniques to prepare arbitrary unbalanced Bell states without
postselection have been demonstrated and applied to tests of
LR~\cite{White, Brida}.

The number of SDs of violation is not normally explicitly converted to
a $p$-value as done here. Instead, it is primarily intended as a way
of claiming successful violation with a good signal-to-noise ratio.
Naturally, one would like to use the measure to compare the strength
of the violation for different experiments. Such a relative comparison
works only if the experiments use the same test of LR with the same
state, experimental settings, losses, visibilities, and other relevant
parameters.  From Fig.~\ref{fig:gain_unbalanced}, we can infer that, if
we use the number of SDs to compare the violation of the CHSH inequality
in experiments involving different unbalanced Bell states, we tend to
unfairly favor the experiment with the more balanced state.

\subsection{Martingale-based Protocol}

Another problem with the SD-based protocol is that it assumes that the
trials are independent and identically distributed; that is, it does
not consider the memory effect~\cite{Barrett2}. We cannot expect the
prepared states and experimental settings to be stable over the course
of a long sequence of trials. In addition, it is desirable to take
into account the possibility that the experimental system is subject
to a model of LR where the entire history of the experiment can affect
the events to come, except that the measurement-setting choices are
still under independent experimental control. To account for these
effects, R. Gill suggested a method for calculating $p$-values based
on the martingale structure of the time sequence of observations in a
test of LR~\cite{Gill1,Gill2}.

The martingale-based $p$-value is computed according to
\begin{equation}
p^{(\text{mart})}=
\text{exp}\left(-\frac{n(\hat{I}-B)^2}{32}\right). \label{confidence2}
\end{equation}
Here, we assume without loss of generality that $I(x)$ and $B$ have
been shifted and normalized so that for every argument $x$, the value
$I(x)$ is bounded between $-4$ and $4$.  If the function $I(x)$ in a
Bell inequality $\langle I(x)\rangle\leq B$ does not satisfy this
condition, then determine $b_l=\min_x I(x)$, $b_u=\max_x I(x)$ and
replace $I(x)$ and $B$ by $I'(x)=8(I(x)-b_l)/(b_u-b_l)-4$ and
$B'=8(B-b_l)/(b_u-b_l)-4$.  The martingale-based protocol is valid,
but is based on conservative tail estimates and therefore is not
asymptotically optimal. For large $n$, $\hat{I}$ approaches $\bar{I}$,
thus the asymptotic confidence-gain rate is
\begin{equation}
G_{\text{mart}} = \log_2(e)\frac{(\bar{I}-B)^2}{32}.
\label{eq:mart_gainrate}
\end{equation}

A comparison of SD-based, martingale-based, and asymptotically optimal
confidence-gain rates is shown in Fig.~\ref{fig:gain_balanced} for a
CHSH test with noisy and lossy Bell states.

\begin{figure}[htb!]
   \includegraphics[scale=0.56, viewport=3cm 9.5cm 22.5cm 19cm]{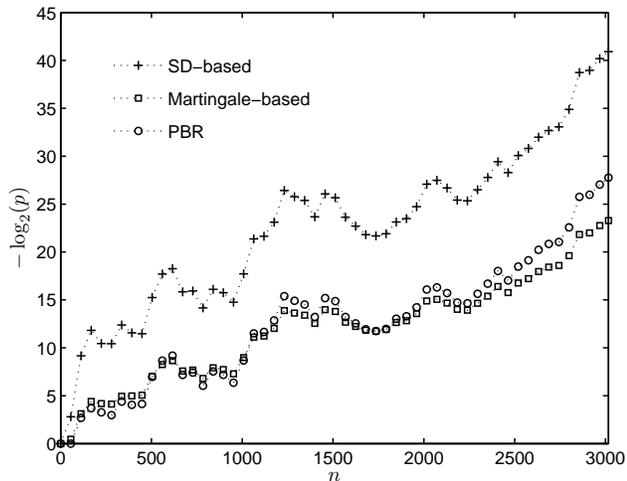}
   \small{
   \caption{Running log-$p$-values as a function of the number of
     trials $n$ in the experiment of Ref.~\cite{Monroe}.  The dotted
     lines are provided only to guide the eye.}
   \label{fig:monroe}}
\end{figure} 

\subsection{PBR Protocol}
\label{subsect:pbr_results}

In contrast to a fixed Bell inequality used in the SD-based or
martingale-based protocol, after $k$ trials but before the $(k+1)$'th
trial the PBR protocol returns a special Bell inequality of the form
\begin{equation}
\langle R_k(x)\rangle \leq 1
\label{eq:PBR_ineqs}
\end{equation}
with $R_k(x)$ nonnegative. The PBR $p$-values are determined
by the values of $R_k$ at the setting and outcome combination
$x_{k+1}$ of the $(k+1)$'th trial. In particular, as shown in
Sec.~\ref{subsect:pbr_theory}, any such sequence of inequalities
yields a valid $p$-value computed according to
\begin{equation}
p^{(\text{PBR})} = \min\left(\left(\prod_{k=1}^n R_{k-1}(x_k)\right)^{-1},1\right).
\label{eq:pbr_confidence_def}
\end{equation}
The PBR protocol aims to optimize the expected $p$-value by computing the PBRs
$R_k(x) = q_{x}^{(k)}/p_{\text{LR},x}^{(k)}$, where $q_{x}^{(k)}$ is
an estimate of the distribution of future setting and outcome
combinations $x$, which can be based on $x_1,\ldots, x_k$ and can take
into account other experimental information obtained before starting
the $(k+1)$'th trial. The quantity in the denominator,
$p_{\text{LR},x}^{(k)}$, is the probability of $x$ given by an optimal
LR model with respect to the estimates $q_{x}^{(k)}$. The notion of
optimality is defined in Sec.~\ref{subsect:pbr_theory} and guarantees
the desired inequality~\eqref{eq:PBR_ineqs}.  We define the (negative)
log-$p$-value increment for the $k$'th trial as
$\log_2(R_{k-1}(x_k))$. For independent and identically distributed trials, $q_{x}^{(k)}$ converges to
the true probabilities $q_{x}$, and the asymptotic confidence-gain
rate is
\begin{equation}
G_{\text{PBR}} = S_q,\label{eq:pbr_gain_rate}
\end{equation}
where $S_q$ is the statistical strength defined in
Sec.~\ref{subsect:pbr_theory}. This is the optimal valid
confidence-gain rate for a given test configuration and is plotted in
Figs.~\ref{fig:gain_unbalanced} and~\ref{fig:gain_balanced}.

\subsection{Application to Experiments}
\label{subsect: application}

The above protocols can compute $p$-values for recorded trials as an
experiment progresses, and such ``running'' $p$-values may be used to
optimize experimental settings.  Because we are interested in
extremely small $p$-values with exponential asymptotic behavior, we
generally consider and display the (negative) log-$p$-value.

SD-based and martingale-based protocols are restricted to a fixed Bell
inequality. The PBR protocol does not have this restriction, which
enables wider searches for strong LR violation. Running log-$p$-values
are shown for a simulation in Fig.~\ref{fig:simulation} and for data
from Ref.~\cite{Monroe} in Fig.~\ref{fig:monroe}.  The PBR $p$-values
were computed with our implementation of the local realism analysis
engine; see the documentation and code. Relevant aspects of the
implementation such as data blocking and learning transients are
discussed in App.~\ref{sect:estimation}.  Note that whereas running
log-$p$-values can be useful for monitoring and tweaking an
experiment, they must not be used as a stopping criterion once an
experiment has been configured.

For Fig.~\ref{fig:simulation} we simulated a CHSH test of LR with a
Bell state and measurement settings maximizing violation of the CHSH
inequality~\eqref{CHSH}. We assumed an ideal experiment (no loss of
photons or visibility) and simulated $5000$ successive trials.
The log-$p$-values were updated for successive blocks of $56$ trials 
(see App.~\ref{sect:estimation}).  In particular, the function $R_k(x)$
used by the PBR protocol was recomputed based on the trials seen so
far every $56$ trials.  The figure shows typical and average runs and
compares the running log-$p$-values to the asymptotic lines with
slopes given by the respective gain rates.  The slopes of the 
running log-$p$-values approach the gain rates, but PBR log-$p$-values
have a systematic offset that can be attributed to an initial 
transient where the setting and outcome distribution is being 
learned. The transient can be removed if, before
the experiment is started, we have a good estimate of the
distribution.  Such an estimate could be based on theory (quantum or
otherwise) or previous measurements, and can be used to ``prime'' the
ratios $R_k(x)$.

For Fig.~\ref{fig:monroe}, we compute log-$p$-values for the data from
the experiment described in Ref.~\cite{Monroe}.  In this experiment,
two $^{171}\text{Yb}^+$ ions separated by about one meter were
entangled through a probabilistic process. In this process, each ion
is entangled with one emitted photon. By projecting the two emitted
photons into a Bell state the two remote ions are entangled with each
other. On the entangled two-ion system, a CHSH test of LR was
performed. The results from $3016$ trials were recorded.  The resulting
estimate of the CHSH expression is $\tilde{I}_{\text{CHSH}}=2.414\pm0.058$.  
For the figure, we processed the data in blocks of $56$ trials as before. 
We did not prime the ratios $R_k(x)$ for computing
PBR log-$p$-values.  In this case, there is insufficient data for 
PBR log-$p$-values to clearly exceed martingale-based ones.

\section{Theory}
\label{sect:theory}
\label{sect:methods}

For SD-based and martingale-based protocols we fix a Bell inequality
$\bar I \leq B$, as explained at the beginning of
Sec.~\ref{sect:results}. While the theory applies to multipartite
Bell inequalities, we discuss it explicitly for the case of
bipartite inequalities to simplify the formulas. (Our implementation
of the local realism analysis engine is presently restricted to the
bipartite case.)  The setting and outcome combination of the $k$'th
trial is denoted by $x_k=(i_k,j_k,a_k,b_k)$, where $i_k$, $j_k$ are the
$k$'th settings and $a_k$, $b_k$ are the $k$'th outcomes of Alice and
Bob, respectively.  Let $i(x)$ and $j(x)$ be Alice's and Bob's
settings, respectively, for the combination $x$.  The distribution of
measurement settings is fixed. The probability of settings $i,j$ is
given by $p_{i,j}$.

\subsection{SD-based Protocol}
\label{subsect:sd_theory}

The obvious method for estimating $\bar{I}$ is to compute the average
of the sequential values $I(x_k)$ given by $\hat I
=\frac{1}{n}\sum_{k=1}^n I(x_k)$. However, this is not the
minimum-variance estimate of $\bar I$, since the setting distribution
is fixed and known.  In fact, the conventional way of writing a Bell
inequality is as a sum of expectations as in Eq.~\eqref{CHSH}, which
makes it independent of the probability distribution of the
settings. The correspondence between the two ways of writing a Bell
inequality is given by
\begin{equation}
\langle I(x)\rangle 
  = \sum_{i,j} p_{i,j}\langle I(x)|i(x)=i,j(x)=j\rangle,
\label{eq:bell_corr}
\end{equation}
where the expectation in the sum is conditioned on the settings of $x$,
as indicated.  If we assume that the state in each trial is
identical and do not worry about the memory and locality loopholes, we
can estimate each expectation $\langle I(x)|i(x)=i,j(x)=j\rangle$
separately, experimentally fixing the settings for each estimate, if
desired.  The right-hand side of Eq.~\eqref{eq:bell_corr} can then be
computed formally.  If we define $n(i,j,a,b)$ to be the number of
trials with settings $i,j$ and outcomes $a,b$, the estimate for
$\bar I$ thus computed is
\begin{equation}
\tilde I = \sum_{i,j} p_{i,j} 
   \frac{\sum_{a,b} n(i,j,a,b) I(i,j,a,b)}{\sum_{a,b} n(i,j,a,b)},
   \label{experimenters' estimate}
\end{equation}
a nonlinear function of $n(i,j,a,b)$. Its SD can be approximated by
linear propagation of errors from SDs for the counts $n(i,j,a,b)$,
assuming each of these counts follows a Poisson distribution. The SD
thus obtained is generally smaller than that of $\hat I$. Hence, the 
conventional way of estimating $\bar I$ and the experimental
SD worsens the validity problem for SD-based $p$-values. However, using 
the estimate $\hat I$ and the associated larger SD in the figures 
of Sec.~\ref{sect:results} does not significantly alter the plots or
their interpretation. 

To convert the number of SDs to a $p$-value, we make the unwarranted
assumption that, for any LR model, the distribution of
the estimate $\tilde I_{\text{LR}}$ of $\bar{I}$ is sufficiently close
to Gaussian with the SD $\sigma$ calculated according to the previous
paragraph but with a mean bounded by $B$. With this assumption,
according to any LR model, the probability of the event $\tilde
I_{\text{LR}} \geq \tilde I$ is then bounded above by $Q((\tilde I -
B)/\sigma)$, which allows us to assign the $p$-value given in
Eq.~\eqref{eq:confidence_sd}, with the caveat that our assumption is
false. The comparisons between SD-based and asymptotically optimal
confidence-gain rates show that this strategy for obtaining $p$-values
is invalid. While it may be possible to obtain a valid $p$-value by
checking the relevant averages and variances for all LR models, this
is a challenging task, and one would still have to consider deviations
from Gaussianity in the extreme tails.

\subsection{Martingale-based Protocol}
\label{subsect:Gill's method}

For fundamental tests of quantum mechanics, a serious deficiency of
SD-based assessments of experimental tests of LR is that they do not
account for memory effects~\cite{Barrett2}, including the possibility
that the state and settings drift in the course of the experiment.
To take such effects into account, R. Gill~\cite{Gill1,Gill2} considered
the time-sequence $M_k = \sum_{l=1}^k(I(x_l)-B)$ as a super-martingale
and applied large-deviation bounds. Here, the measurement settings are
assumed to be chosen randomly and independently by Alice and Bob
according to the fixed probability distribution $p_{i,j}$ built into the
inequality of Eq.~\eqref{eq:proper_Bell}. Let $W_k$ be all the
information available before the $k$'th trial.  According to any LR
model, the conditional expectation of $M_k$ given $W_k$ satisfies
\begin{align}
\langle M_k|W_k\rangle&=\langle I(x_k)-B+M_{k-1}|W_k\rangle \notag \\
            &=\langle I(x_k)|W_k\rangle-B+\langle M_{k-1}|W_k\rangle \notag \\
            &=\langle I(x_k)|W_k\rangle-B + M_{k-1}\notag \\
            &\leq M_{k-1}. \label{martingale}
\end{align}
The last inequality follows from the fact that the Bell
inequality~\eqref{eq:proper_Bell} is satisfied for any LR model,
regardless of prior information.  The inequality in
Eq.~\eqref{martingale} is the defining property for a super-martingale
$\{M_k : k=1,2,\ldots\}$. This inequality is still satisfied if
$I(x)$ and $B$ have been normalized and shifted by some constants so
that $-4\leq I(x)\leq 4$.  With this normalization and shift, each
``increment'' $M_k-M_{k-1}$ of the super-martingale is bounded between
$b_l=-4-B$ and $b_u=4-B$.  By applying the Azuma-Hoeffding
inequality~\cite{Hoeffding, Azuma, mcdiarmid:qc1989a}, we find that,
after $n$ trials, the probability that an LR model yields an estimate
$\hat{I}_{\text{LR}}$ greater than or equal to the observed $\hat{I}$
is bounded above by
\begin{align}
\text{Prob}_{\text{LR}}(\hat{I}_{\text{LR}} \geq\hat{I})
&=\text{Prob}_{\text{LR}}(M_n\geq n(\hat{I}-B))
\notag\\ &\leq
\text{exp}\left(-\frac{2n(\hat{I}-B)^2}{(b_u-b_l)^2}\right). 
\label{martingale_bound}
\end{align}
This implies a valid $p$-value of
\begin{equation}
p^{(\text{mart})}=\text{exp}\left(-\frac{2n(\hat{I}-B)^2}{(b_u-b_l)^2}\right). \label{con2} 
\end{equation}
Substituting $b_u-b_l=8$ gives Eq.~\eqref{confidence2}. Note that, for
the CHSH inequality, the expression for martingale-based $p$-values
obtained above improve the expression in Ref.~\cite{Gill2} and the expression
applied to experimental data in Ref.~\cite{Monroe} by taking advantage
of the bounds on $I(x)$ in the formulation of the Azuma-Hoeffding
inequality used here.

We cannot expect the bound on the tail probability in
Eq.~\eqref{martingale_bound} to be asymptotically optimal, since the
only constraints considered are the bounds of $I(x)$.  The PBR
protocol takes advantage of all available constraints on the setting
and outcome distributions, implicitly including all relevant Bell
inequalities.

\subsection{PBR Protocol}
\label{subsect:pbr_theory}

Let $R_k(x)$, $k=0,1,\ldots$ be a sequence of PBRs as introduced in
Sec.~\ref{subsect:pbr_results}.  They are designed so that $0\leq
R_k(x)$ and $\langle R_k(x)\rangle\leq 1$ for any LR model, provided
that the setting distribution is $p_{i,j}$.  Here, $R_k$ may depend on
$x_1,\ldots,x_k$ and other aspects of the experiment before starting
the $(k+1)$'th trial.  We now show that any sequence of $R_k$ with
these properties satisfies that the $p$-value computed according to
Eq.~\eqref{eq:pbr_confidence_def} is valid.

As in the previous section, we let $W_k$ denote all the information
available before the $k$'th trial. Let $P_k=\prod_{l=1}^k
R_{l-1}(x_l)$.  According to any LR model with arbitrary memory, the
expectation of $P_k$ conditioned on $W_k$ satisfies
\begin{align}
\langle P_{k}|W_k \rangle 
  &= \left\langle \prod_{l=1}^{k}R_{l-1}(x_l)|W_k \right\rangle \notag\\ 
  &= \left\langle \prod_{l=1}^{k-1}R_{l-1}(x_l) \times R_{k-1}(x_{k})|W_k \right\rangle \notag\\ 
  &= \prod_{l=1}^{k-1} R_{l-1}(x_l) \times \langle R_{k-1}(x_{k})| W_k \rangle \notag\\
  &\leq P_{k-1},
\label{eq:P_n}
\end{align}
where we used the facts that $W_k$ includes $R_{l-1}$ and $x_{l-1}$
for $l\leq k$, and that the LR bound on $\langle R_{k-1}(x)\rangle$ is
$1$ given $W_k$, as the LR model in the bound is arbitrary.  We can
compute the expectations of both sides of Eq.~\eqref{eq:P_n} to show
that, according to any LR model, $\langle P_{k}\rangle \leq \langle
P_{k-1}\rangle$, and therefore, by induction, $\langle P_k\rangle\leq
1$.

Given a sequence of experimental results $x_1,\ldots, x_n$ from $n$
trials, the random variable $P_n$ takes a specific value $\hat P$.
Suppose that $P_n$ is constrained by LR, possibly with
memory. By construction, $P_n\geq 0$ and the expectation of $P_n$
according to this model is bounded above by $1$. According to Markov's
inequality, we conclude that
\begin{equation}
\text{Prob}_{\text{LR}}(P_n \geq \hat P) \leq \min(1/\hat P,1),
\end{equation}
which shows that we can assign a valid $p$-value for rejecting LR by
setting $p^{(\text{PBR})} = \min(1/\hat P,1)$ as in
Eq.~\eqref{eq:pbr_confidence_def}.  Note that Eq.~\eqref{eq:P_n} shows
that the sequence $P_k$, $k=1,2,\ldots$ is a super-martingale under
any LR model. However, this super-martingale's ``increment'' is not
bounded, so we cannot directly apply the method of
Sec.~\ref{subsect:Gill's method} to bound the tail probability.

For the extremely low $p$-values of interest in tests of LR, we are
looking for large log-$p$-value increments $\log_2(R_{n}(x_{n+1}))$ at
the $(n+1)$'th trial.  Therefore, before the $(n+1)$'th trial, our
goal is to choose $R_{n}(x)$ so as to maximize the experimentally
expected increment $\langle\log_2(R_{n}(x_{n+1}))\rangle$.  For this
purpose, we can take advantage of anything we know about the
probability distribution of the result $x_{n+1}$ to be obtained at the
next trial.  Consider a probability distribution $q$ for $x_{n+1}$,
which may be either the true distribution or an estimate thereof.  Let
$p$ be the distribution according to an LR model. Note that, because
the setting distribution is under experimental control, the
probability distributions $q$ and $p$ must be consistent with the
chosen setting distribution.  Our ability to distinguish the
probability distributions $q$ and $p$ given a collection of
independent samples from $q$ can be characterized by the
asymptotically optimal confidence-gain rate for rejecting $p$ in favor
of $q$. As shown in Ref.~\cite{Bahadur}, this optimal rate is given by
the Kullback-Leibler (KL) divergence from $q$ to $p$,
\begin{equation}
D_{\text{KL}}(q|p)= \sum_x q_x \log_2(q_x/p_x).
\label{KL divergence}
\end{equation}
The KL divergence is nonnegative, and it is zero iff $p=q$.  This
motivates seeking an LR model whose probability distribution
$p_{\text{LR}}$ minimizes the KL divergence from $q$~\cite{VanDam}.
This is the optimal LR model mentioned in
Sec.~\ref{subsect:pbr_results}.  We define $S_q =
D_{\text{KL}}(q|p_{\text{LR}})$, and refer to $S_q$ as the statistical
strength for rejecting LR by means of a test with the distribution
$q$. 

We claim that if we define $R_n(x) = q_x/p_{\text{LR,x}}$, then $0\leq
R_n(x)$, and for any LR model, the expectation satisfies $\langle
R_n(x)\rangle\leq 1$. Consequently, the $p$-value computed according
to Eq.~\eqref{eq:pbr_confidence_def} is valid.  To prove the claim,
consider
$\phi(\beta)=D_{\text{KL}}(q|p_{\text{LR}}+\beta(p-p_{\text{LR}}))$,
where $0\leq \beta \leq 1$. For any $p$ in the convex set of LR
distributions, by optimality of $p_{\text{LR}}$,
$\phi(\beta)\geq\phi(0)$. It follows that $\frac{\partial
  \phi}{\partial \beta}|_{\beta=0+}\geq 0$. Consequently,
\begin{equation}
\sum_x (p_{\text{LR},x}-p_x)\frac{q_x}{p_{\text{LR},x}} \geq 0,
\end{equation}
which can be rearranged to show that
\begin{equation}
\langle R_n(x)\rangle_p = \sum_x p_x \frac{q_x}{p_{\text{LR,x}}} \leq 1.
\label{eq:PBR_bell_ineq}
\end{equation}
The claim follows.  Bell inequalities of the form shown in
Eq.~\eqref{eq:PBR_bell_ineq}, which are based on minimizing the 
KL divergence, were introduced in Ref.~\cite{Acin}.

Consider the choice $R_n(x) = q_x/p_{\text{LR,x}}$ made before the
$(n+1)$'th trial. If $q$ is the true distribution of $x_{n+1}$, then
the experimental expectation $\bar l=\langle
\log_2(R_{n}(x_{n+1}))\rangle$ is the statistical strength
$S_q$. Since $\bar l$ is the expected log-$p$-value increment, which
cannot exceed $S_q$~\cite{Bahadur}, this choice of $R_n$ maximizes the
confidence-gain rate. However, we do not know the true distribution
$q$.  Instead, we obtain good estimates $q'$ of $q$ before the
$(n+1)$'th trial, and determine the corresponding optimal LR model's
probability distribution $p'_{\text{LR}}$. We then set $R_n(x) =
q'_x/p'_{\text{LR,x}}$ to compute and update the PBR $p$-value.  If
the experiment is sufficiently stable, good estimates can be obtained
from the frequencies of events observed in trials so far. The
estimates can be improved by taking into account that the setting
distribution is known and the distributions of marginal outcomes for
given settings of Alice or Bob must agree due to the no-signaling
constraints. We discuss how to do this in App.~\ref{sect:estimation}.
In App.~\ref{sect:bad estimation}, we show that if the trials are
independent and identically distributed, then PBR $p$-values 
computed with any converging method for
estimating the true setting and outcome distribution $q$ have the
property that the confidence-gain rate approaches the statistical
strength $S_q$, thus proving asymptotic optimality of PBR $p$-values.

To determine the optimal LR model one can use numerical algorithms for
optimizing convex functions on a convex domain.  In this case one can
use the expectation-maximization (EM) algorithm~\cite{Vardi} as
discussed in~\cite{Zhang}. A problem is that due to stopping criteria
and numerical precision, one cannot expect to find the exact optimum.
We show in App.~\ref{sect:bad estimation} that one can compensate for
this problem to maintain validity of the computed $p$-value.

\section{Conclusion}
\label{sect:conclusion}

The degree of violation of LR in a Bell-type test is usually expressed
in terms of the number of SDs of violation. This quantity cannot,
however, be used to obtain valid $p$-values for rejecting LR by
conventional means.  It also fails to quantitatively compare the
success of different experimental tests of LR and does not account for
stability issues or memory effects in experiments. We solve these
problems by providing a method---the PBR protocol---for determining
valid $p$-values directly from the settings and outcomes in a sequence
of trials. The PBR protocol does not rely on a predetermined Bell
inequality, adapts to the actual experimental configuration, and is
asymptotically optimal for independent and identically distributed
trials.  It therefore provides a standardized measure of success for
experimental tests of LR. While the protocol remains valid if the
experiment drifts over the sequence of trials, how well it performs
depends on the nature of the drifts and how the protocol takes them
into account.  Another valid protocol that accounts for memory effects
can be based on martingale bounds~\cite{Gill1,Gill2}. This protocol
requires a Bell inequality that is fixed for the experiment. Given the
Bell inequality, the martingale-based protocol has the advantage that
it is computationally efficient with respect to number of settings,
outcomes, and parties. The disadvantage is that it is suboptimal and
does not provide a clear quantitative comparison of different
experimental tests. Our simulations show that it is practical to apply the
PBR protocol to data from typical experimental configurations, and
that the running $p$-values can be used for tweaking an experiment in
progress to find the experimentally accessible configuration that
provides the highest violation of LR.

\begin{acknowledgments}
We thank D. N. Matsukevich for providing the experimental data for
Ref.~\cite{Monroe}, and we are grateful for the editorial support of
A. Migdall and K. Coakley. This paper is a contribution of the
National Institute of Standards and Technology and is not subject to
U.S. copyright.
\end{acknowledgments}

\appendix*
\section{}
\label{sect:appendix}

\subsection{Statistical Concepts}
\label{sect:statistics_terms}

A main purpose of the PBR and related protocols is to evaluate the
strength of the evidence against LR by computing valid $p$-values
given the data. Some care must be taken in interpreting such
$p$-values in terms of probabilities. For example, the $p$-value
\emph{cannot} be interpreted as a probability that LR is
true. Although they are computed for the data, their validity is
defined in terms of what is known before the experiment, not
after. Strictly speaking, we can only state for sure that
\emph{before} performing the trials, the following holds: For any
fixed $0\leq \alpha\leq 1$, if LR holds, then the probability that the
returned $p$-value satisfies $p\leq\alpha$ is at most
$\alpha$. Although we have no intention of making an actual decision
on the failure of LR, this statement can be viewed in terms of
traditional hypothesis testing: The protocol tests LR simultaneously
at all significance levels $\alpha$, and ``rejects'' LR at a given
$\alpha$ if $p\leq\alpha$. The validity property is equivalent to the
statement that, if LR holds, the maximum probability of (falsely)
rejecting at level $\alpha$ is bounded above by $\alpha$. This
justifies the use of $p$-values to quantify LR violation. The
definitions of significance levels and $p$-values are based on
Ref.~\cite{Shao}, $2$nd edition, pages~$126$ and~$127$.

The $p$-values returned by the protocols considered here are defined
in terms of bounds on the one-sided tail probabilities of a test
statistic $T$. For given $T$, it is conventional to define \emph{the}
$p$-value of $T$ given data $\mathbf{x}$ as the supremum of the
tail-probabilities $\text{Prob}(T\geq T(\mathbf{x}))$ over all
hypotheses to be rejected (the null hypotheses). While such tight
$p$-values are desirable, they are impractical to compute in general.
Hence our definition of valid $p$-values requires only an upper bound.
Note that, for our situation, the computation of tight $p$-values is
further complicated by the fact that the set of null hypotheses
includes all possible sequences of LR models depending on previous
trials.  Furthermore, while the statistic is well-defined for any
realization of the PBR protocol, it is not unique.

We use the the term ``protocol'' rather than ``test'' for two
reasons. The first is that the term ``test'' in ``test of LR''
typically refers to the experimental setup and subsequent analysis,
not a conventional hypothesis test. The second is that hypothesis
tests, as the term is used in mathematical statistics, are valid by
definition.  Thus, although we do not encourage it, one can think of a
valid analysis protocol as a family of hypothesis tests. For such a
family to be useful, the tests should also have high power. For our
situation, one can express the power in terms of the probabilities of
rejection at given significance levels and non-LR
models. Alternatively, one can consider the expected $p$-values, and
look for tests for which they are as small as possible. We do not
expect that the PBR protocol has particularly low $p$-values for a
given finite number of trials. In fact, because of
the conservative nature of the Markov bounds, better tests exist.
However, asymptotic optimality of the PBR protocol assures us that it
performs well when the evidence for rejection is very strong.  It is
also worth noting that many issues that arise in applications of
hypothesis testing, such as selection biases, are less of a concern
when one is considering the extremely low $p$-values that are
desirable when falsifying a physical theory. Corrections for such
effects reduce log-$p$-values by relatively small terms in our
setting. Also, one application of the PBR protocol is to quantify the
success of an experiment independent of the details of the
configuration, so that different experiments can be compared. For this
application, the statistical interpretation of the $p$-value
serves only as motivation.

Probability ratios such as the ones we use to compute the values of
$R_k(x)$ in Eq.~\eqref{eq:PBR_ineqs} are often referred to as
likelihood ratios. Likelihood ratios play an important role in many
statistical tests as explained in statistics textbooks such as
Ref.~\cite{Shao}. In the PBR protocol, the statistic 
can be computed from any sequence of nonnegative functions $R_k(x)$ 
satisfying the inequality in Eq.~\eqref{eq:PBR_ineqs}. Thus, the 
probability ratios are simply an intermediate step to obtaining
such functions. We do not ascribe any other meaning to the ratios.

\subsection{Estimating the Setting and Outcome Distribution}
\label{sect:estimation}

Consider $n$ trials with settings and outcomes given by $x_1,\ldots
x_{n}$. Our goal is to obtain an estimate $q'$ of the true probability
distribution $q$ of the $(n+1)$'th trial's settings and
outcomes. Assuming no other knowledge, the estimate can be based on
the empirical frequencies $f_x = \frac{1}{n}\sum_{k=1}^{n}\delta_{x_k,
  x}$.  Due to statistical fluctuations, the empirical frequencies are
not likely to satisfy the following known constraints satisfied by $q$:
\begin{itemize}
\item Setting distribution: The setting distribution $p_{i,j}$ is fixed,
   and $q$ satisfies $\sum_{a,b} q_{(i,j,a,b)}=p_{i,j}$.
\item No-signaling: Given that Alice uses setting $i$, the
  distribution of Alice's measurement outcomes does not depend on
  Bob's settings, and vice versa.
\end{itemize}
There are two other issues for calculating PBR $p$-values.  The first
is that some empirical frequencies $f_x$ may be zero. If our estimate
is $q'=f$, zero frequencies can be disastrous. In the case where the
corresponding settings and outcomes occur in the next trial, the ratio
contributing to the PBR $p$-value in Eq.~\eqref{eq:pbr_confidence_def} 
can be zero, and then the $p$-value goes to $1$ with no possibility 
of later recovery.  The second and related issue is that in the absence
of prior knowledge, initially we have insufficient information to 
make useful estimates of probability distributions of future 
settings and outcomes. Even if the problem of zero frequencies has 
been taken care of, this can still result in initial ``learning'' 
transients that result in a negative offset in the 
accumulated log-$p$-values.

Our approach for estimating the next trial's setting and outcome
distribution uses maximum likelihood to obtain an estimate that
respects the above constraints and then adjusts the estimate by mixing
in a distribution that is uniform conditional on the settings.  To
reduce the impact of learning transients, we process the trials in
blocks.

To apply maximum likelihood for computing a first estimate $q_0$ of $q$, we
assume independent and identically distributed trials.  The probability of observing empirical
frequencies $f$ after $n$ trials given that the true distribution 
is $q$ is proportional to
\begin{equation}
L(f|q) = \prod_x q_{x}^{nf_x}.
\end{equation}
We therefore set $q_0$ according to
\begin{equation}
q_0 = \text{argmax}_{q'\in V} L(f|q'),
\label{eq:ML_pred}
\end{equation}
where $V$ is the set of probability distributions satisfying the
setting distribution and no-signaling constraints.  These constraints
are linear and $\log(L(f|q))$ is concave, so there is no difficulty in
applying available nonlinear optimization tools.  Note that, for the
purpose of calculating PBR $p$-values, it is not critical that
Eq.~\eqref{eq:ML_pred} is exactly satisfied, so it is not necessary to
use extremely tight stopping criteria to ensure identity with the best
numerical precision possible.  Also, whereas the assumptions
underlying Eq.~\eqref{eq:PBR_bell_ineq} require that the setting
distribution constraint is satisfied, the no-signaling constraint is
not critical. Applying it helps improve our estimates, but the effect
on the log-$p$-value increments becomes negligible for large $n$.

There are different ways to solve the problem with empirical
frequencies that are zero; some are explained in
Refs.~\cite{Ristad,Blume-Kohout}. They generally involve mixing in a
distribution that has no zero probabilities with a weight 
that decreases to zero as $n$ grows. For the plots in
Figs.~\ref{fig:simulation} and~\ref{fig:monroe}, we modified $q_0$ by
setting $q_1=\frac{n}{n+1} q_0 + \frac{1}{n+1} u$, where conditionally
on the settings, $u$ is uniform, and $u$'s setting distribution is
$p_{i,j}$.

There are different approaches to mitigating the effect of the initial
learning transient. The first is to ``prime'' the estimates with
knowledge about the experiment available before the trials are
started. Such knowledge could be based on theory or on experiments
designed to characterize the quantum state and measurement setup.  The
prior information must be assigned a weight. In our implementation of
the local realism analysis engine, the weight is determined by the
number of trials that would have been required to obtain an equally
good estimate directly from the frequencies. Proper use of priming
requires that the initial estimates and parameters such as the weight
are determined ``blindly'' before any knowledge of the actual data to
be analyzed is available.

A second approach is to set $R_{n}(x)=1$ unless the statistical
strength $S$ for $q_1$'s violation of LR seems sufficiently
significant given that the estimated distribution $q_1$ is based on
$n$ trials. While one might expect that the violation is sufficiently
significant if $nS\geq c$ for some constant $c$, simulations show that
the best choice of $c$ depends on the distribution of settings and
outcomes in the experiment.

The third and simplest approach is to block the data from the trials.
Instead of updating the log-$p$-value after every trial, we process
data $h$ trials at a time. The first block is used only for estimating
the setting and outcome distribution of future trials. That is, we set
$R_k(x)=1$ for $k=0,\ldots,(h-1)$. Subsequently, we have
$R_{mh+k}=R_{mh}$ for $k=1,\ldots,(h-1)$ and all $m$. Note that
neither the validity nor the asymptotic optimality of the calculated
$p$-values requires updating the PBRs after each trial. Choosing $h$
large enough ensures that the first block's trials have sufficient
information for obtaining reasonable estimates of the distribution.
An additional advantage of blocking the trials is that we avoid
unnecessarily invoking the computationally costly optimizations
required for updating the PBRs. We standardized the choice of block
size so that if the total number of trials to be analyzed is $N$, $h$
is the maximum of $\lceil N/1000\rceil$ and $\lceil\ln(2d)d\rceil$,
where $d$ is the number of possible setting and outcome combinations
in a trial. The first expression ensures that we do not lose
too much log-$p$-value by using the first block only for learning 
the setting and outcome distribution. The second one is chosen 
so that if $q$ is uniform, the probability that every setting 
and outcome combination occurs is at least $1/2$.

We conclude this section with a note on implementing the PBR protocol.
For monitoring an experiment and to adapt to changes in experimental
configuration, the estimated setting and outcome distributions used in
the PBRs should be based on recent trials only. This can be
accomplished by windowing the trials with a window large enough to
have statistically significant violation of LR (if there is
violation), but small enough to avoid seeing significant changes in
configuration. Our implementation of the local realism analysis engine
uses a computationally simpler approach based on weighting the trials
with exponentially decreasing weights in time determined by a
configurable half-life. This feature was not used in the comparisons
in Sec.~\ref{sect:results}.

\subsection{Effects of Suboptimal Estimates and LR Models}
\label{sect:bad estimation}

Ideally the estimated distribution $q'$ used in the numerator of
$R_{n}$ matches the true distribution $q$, and the LR distribution
$p'_{\text{LR}}$ in the denominator of $R_{n}$ exactly minimizes the
KL divergence from $q'$. As shown in Sec.~\ref{subsect:pbr_theory}, 
having $q'$ different from $q$ does not affect the validity of the 
PBR $p$-values. But it can reduce the expected log-$p$-value 
increment $\overline{l}$.  Let $S_{q}$ be the statistical strength 
of $q$ for LR violation. We show that
\begin{equation}
S_{q} \geq \overline{l} \geq S_{q}-D_{\text{KL}}(q|q').
\label{eq:slp0}
\end{equation}
For reasonable methods of estimating $q'$ such as the one described
in App.~\ref{sect:estimation} and independent and identically distributed trials, $q'$ almost surely
approaches $q$ so that $D_{\text{KL}}(q|q')$ goes to zero. This
shows that the PBR protocol has asymptotic confidence-gain rate $S_{q}$.

To prove the first inequality in Eq.~\eqref{eq:slp0}, let
$p_{\text{LR}}$ be the LR distribution that minimizes the
KL divergence from $q$, so that $S_q=D_{\text{KL}}(q|p_{\text{LR}})$.
We bound $\overline{l}$ as follows:
\begin{eqnarray}
S_q-\overline{l} &=&\sum_x  q_{x}\log_2(q_{x}/p_{\text{LR},x}) 
        -\sum_x q_{x} \log_2(q'_x/p'_{\text{LR},x}) \notag\\
  &=& \sum_x  q_{x}\log_2(q_{x}/t_x),
        \label{eq:slp1}
\end{eqnarray}
where we define $t_x = p_{\text{LR},x} q'_x/p'_{\text{LR},x}$.
Since $q'_x/p'_{\text{LR},x}$ is the PBR, and $p_{\text{LR}}$ is
an LR distribution, we know that
$c\equiv\sum_x t_x \leq 1$ [Eq.~\eqref{eq:PBR_bell_ineq}].
Since $t'=t/c$ is a probability distribution, we can continue
the calculation:
\begin{equation}
S_q-\overline{l}= \log_2(1/c)+\sum_x q_{x}\log_2(q_{x}/t'_x)\geq 0,\label{eq:slp2}
\end{equation}
because the second term is a KL divergence. 

To obtain the second inequality of Eq.~\eqref{eq:slp0} we bound
\begin{eqnarray}
\overline{l} 
  &=& \sum_x q_{x} \log_2(q'_x/p'_{\text{LR},x}) \notag\\
  &=& \sum_x q_{x} \log_2(q_{x}/p'_{\text{LR},x})
      - \sum_x q_{x}\log_2(q_{x}/q'_x) \notag \\
  &=& D_{\text{KL}}(q|p'_{\text{LR}}) - D_{\text{KL}}(q|q') \notag \\
  &\geq& D_{\text{KL}}(q|p_{\text{LR}}) - D_{\text{KL}}(q|q') \notag \\
  &=& S_q-D_{\text{KL}}(q|q').
\end{eqnarray}

The denominator $p'_{\text{LR}}$ of the PBRs $R_{n}$ must be computed
numerically. Consequently, the distribution $p'_{e,\text{LR}}$
actually obtained is typically not identical to $p'_{\text{LR}}$ and
may not minimize the relevant KL divergence. Hence, there may be an LR
distribution $p$, for which $\langle R'_{n}(x)\rangle_p = \langle
q'_x/p'_{e,\text{LR},x}\rangle_p$ is greater than $1$, and so the PBR
$p$-value is not valid if it is computed according to 
Eq.~\eqref{eq:pbr_confidence_def} with $R'_n$. To maintain validity, 
we determine the maximum value $1+\epsilon$ of $\langle
R'_{n}(x)\rangle_p$ for all LR distributions $p$ and then set $R_n
=R'_n/(1+\epsilon)$.  To determine the bound
$1+\epsilon$, we recall that LR distributions are mixtures of
distributions $p_{\lambda}$ induced by ``local hidden variables''
$\lambda$. Each $\lambda$ assigns deterministic outcomes independently
for each setting of Alice and each setting of Bob. We write
$\lambda_{A,i}$ and $\lambda_{B,j}$ for Alice's and Bob's measurement
outcomes given settings $i$ and $j$, according to $\lambda$. The
probability for the setting and outcome combination $x=(i,j,a,b)$ is
given by $p_{\lambda,(i,j,a,b)} =
p_{i,j}\delta_{a,\lambda_{A,i}}\delta_{b,\lambda_{B,j}}$.  With these
definitions,
\begin{equation}
1+\epsilon = \max_{\text{$p$ is LR}}\langle q'_x/p'_{e,\text{LR},x}\rangle_p =
\max_{\lambda} \sum_x p_{\lambda, x}
q'_x/p'_{e,\text{LR},x}.\label{eq:pbr_bound}
\end{equation}
Because the number of different $\lambda$ is finite, the value
$1+\epsilon$ can be calculated according to Eq.~\eqref{eq:pbr_bound}.
The EM algorithm that we apply to KL-divergence minimization
iteratively updates the probability distribution over the set of
hidden variable assignments $\lambda$. To perform the updates requires
the set of values that are maximized in Eq.~\eqref{eq:pbr_bound}, so
the computation of $1+\epsilon$ can be integrated into the algorithm
with little overhead.  Furthermore, the quantity $\epsilon$ can be
used as a stopping criterion for minimization.  That is, the expected
log-$p$-value increment $\bar l_e$, assuming that the result $x$ is
distributed according to $q'$, satisfies
\begin{eqnarray}
\bar l_e &=& \sum_x q'_x \log_2(q'_x/(p'_{e,\text{LR},x}(1+\epsilon))) \notag\\
  &=& D_{\text{KL}}(q'|p'_{e,\text{LR}}) - \log_2(1+\epsilon) \notag\\
  &\geq& D_{\text{KL}}(q'|p'_{\text{LR}}) - \log_2(1+\epsilon).
\end{eqnarray}
Thus, for independent and identically distributed trials, the
asymptotic confidence-gain rate is lowered by at most
$\log_2(1+\epsilon)$.

\bibliography{prospects}

\begin{thebibliography}{29}
\expandafter\ifx\csname natexlab\endcsname\relax\def\natexlab#1{#1}\fi
\expandafter\ifx\csname bibnamefont\endcsname\relax
  \def\bibnamefont#1{#1}\fi
\expandafter\ifx\csname bibfnamefont\endcsname\relax
  \def\bibfnamefont#1{#1}\fi
\expandafter\ifx\csname citenamefont\endcsname\relax
  \def\citenamefont#1{#1}\fi
\expandafter\ifx\csname url\endcsname\relax
  \def\url#1{\texttt{#1}}\fi
\expandafter\ifx\csname urlprefix\endcsname\relax\def\urlprefix{URL }\fi
\providecommand{\bibinfo}[2]{#2}
\providecommand{\eprint}[2][]{\url{#2}}

\bibitem[{\citenamefont{Bell}(1964)}]{Bell}
\bibinfo{author}{\bibfnamefont{J.~S.} \bibnamefont{Bell}},
  \bibinfo{journal}{Physics} \textbf{\bibinfo{volume}{1}}, \bibinfo{pages}{195}
  (\bibinfo{year}{1964}).

\bibitem[{\citenamefont{Clauser et~al.}(1969)\citenamefont{Clauser, Horne,
  Shimony, and Holt}}]{Clauser}
\bibinfo{author}{\bibfnamefont{J.~F.} \bibnamefont{Clauser}},
  \bibinfo{author}{\bibfnamefont{M.~A.} \bibnamefont{Horne}},
  \bibinfo{author}{\bibfnamefont{A.}~\bibnamefont{Shimony}}, \bibnamefont{and}
  \bibinfo{author}{\bibfnamefont{R.~A.} \bibnamefont{Holt}},
  \bibinfo{journal}{Phys. Rev. Lett.} \textbf{\bibinfo{volume}{23}},
  \bibinfo{pages}{880} (\bibinfo{year}{1969}).

\bibitem[{\citenamefont{Weihs et~al.}(1998)\citenamefont{Weihs, Jennewein,
  Simon, Weinfurter, and Zeilinger}}]{Weihs}
\bibinfo{author}{\bibfnamefont{G.}~\bibnamefont{Weihs}},
  \bibinfo{author}{\bibfnamefont{T.}~\bibnamefont{Jennewein}},
  \bibinfo{author}{\bibfnamefont{C.}~\bibnamefont{Simon}},
  \bibinfo{author}{\bibfnamefont{H.}~\bibnamefont{Weinfurter}},
  \bibnamefont{and}
  \bibinfo{author}{\bibfnamefont{A.}~\bibnamefont{Zeilinger}},
  \bibinfo{journal}{Phys. Rev. Lett.} \textbf{\bibinfo{volume}{81}},
  \bibinfo{pages}{5039} (\bibinfo{year}{1998}), \bibinfo{note}{the statistical
  error is rounded at the last digit. See Weihs' PhD thesis online
  (\url{http://old.iqc.uwaterloo.ca/~gweihs/gwdiss.pdf}) for details.}

\bibitem[{\citenamefont{Shao}(2003)}]{Shao}
\bibinfo{author}{\bibfnamefont{J.}~\bibnamefont{Shao}},
  \emph{\bibinfo{title}{Mathematical Statistics}}
  (\bibinfo{publisher}{Springer, New York}, \bibinfo{year}{2003}),
  \bibinfo{edition}{2nd} ed.

\bibitem[{\citenamefont{Gill}(2003{\natexlab{a}})}]{Gill1}
\bibinfo{author}{\bibfnamefont{R.~D.} \bibnamefont{Gill}}, in
  \emph{\bibinfo{booktitle}{Mathematical Statistics and Applications:
  Festschrift for Constance van Eeden. Eds: M. Moore, S. Froda and C. L\'eger.
  IMS Lecture Notes -- Monograph Series}} (\bibinfo{publisher}{Institute of
  Mathematical Statistics. Beachwood, Ohio},
  \bibinfo{year}{2003}{\natexlab{a}}), vol.~\bibinfo{volume}{42}, pp.
  \bibinfo{pages}{133--154}, \bibinfo{note}{also available as
  arXiv:quant-ph/0110137.}

\bibitem[{\citenamefont{Gill}(2003{\natexlab{b}})}]{Gill2}
\bibinfo{author}{\bibfnamefont{R.~D.} \bibnamefont{Gill}}, in
  \emph{\bibinfo{booktitle}{Proc. of ``Foundations of Probability and Physics -
  2", Ser. Math. Modelling in Phys., Engin., and Cogn. Sc.}}
  (\bibinfo{publisher}{V\"axj\"o Univ. Press.},
  \bibinfo{year}{2003}{\natexlab{b}}), vol.~\bibinfo{volume}{5}, pp.
  \bibinfo{pages}{179--206}.

\bibitem[{\citenamefont{van Dam et~al.}(2005)\citenamefont{van Dam, Gill, and
  Grunwald}}]{VanDam}
\bibinfo{author}{\bibfnamefont{W.}~\bibnamefont{van Dam}},
  \bibinfo{author}{\bibfnamefont{R.~D.} \bibnamefont{Gill}}, \bibnamefont{and}
  \bibinfo{author}{\bibfnamefont{P.~D.} \bibnamefont{Grunwald}},
  \bibinfo{journal}{IEEE Trans. Inf. Theory} \textbf{\bibinfo{volume}{51}},
  \bibinfo{pages}{2812} (\bibinfo{year}{2005}).

\bibitem[{\citenamefont{Barrett et~al.}(2002)\citenamefont{Barrett, Collins,
  Hardy, Kent, and Popescu}}]{Barrett2}
\bibinfo{author}{\bibfnamefont{J.}~\bibnamefont{Barrett}},
  \bibinfo{author}{\bibfnamefont{D.}~\bibnamefont{Collins}},
  \bibinfo{author}{\bibfnamefont{L.}~\bibnamefont{Hardy}},
  \bibinfo{author}{\bibfnamefont{A.}~\bibnamefont{Kent}}, \bibnamefont{and}
  \bibinfo{author}{\bibfnamefont{S.}~\bibnamefont{Popescu}},
  \bibinfo{journal}{Phys. Rev. A} \textbf{\bibinfo{volume}{66}},
  \bibinfo{pages}{042111} (\bibinfo{year}{2002}).

\bibitem[{\citenamefont{Ac\'in et~al.}(2007)\citenamefont{Ac\'in, Brunner,
  Gisin, Massar, Pironio, and Scarani}}]{Acin2}
\bibinfo{author}{\bibfnamefont{A.}~\bibnamefont{Ac\'in}},
  \bibinfo{author}{\bibfnamefont{N.}~\bibnamefont{Brunner}},
  \bibinfo{author}{\bibfnamefont{N.}~\bibnamefont{Gisin}},
  \bibinfo{author}{\bibfnamefont{S.}~\bibnamefont{Massar}},
  \bibinfo{author}{\bibfnamefont{S.}~\bibnamefont{Pironio}}, \bibnamefont{and}
  \bibinfo{author}{\bibfnamefont{V.}~\bibnamefont{Scarani}},
  \bibinfo{journal}{Phys. Rev. Lett.} \textbf{\bibinfo{volume}{98}},
  \bibinfo{pages}{230501} (\bibinfo{year}{2007}).

\bibitem[{\citenamefont{Masanes}(2009)}]{Masanes}
\bibinfo{author}{\bibfnamefont{L.}~\bibnamefont{Masanes}},
  \bibinfo{journal}{Phys. Rev. Lett.} \textbf{\bibinfo{volume}{102}},
  \bibinfo{pages}{140501} (\bibinfo{year}{2009}).

\bibitem[{\citenamefont{Pironio\emph{ et al.}}(2010)}]{Monroe}
\bibinfo{author}{\bibfnamefont{S.}~\bibnamefont{Pironio\emph{ et al.}}},
  \bibinfo{journal}{Nature} \textbf{\bibinfo{volume}{464}},
  \bibinfo{pages}{1021} (\bibinfo{year}{2010}).

\bibitem[{\citenamefont{Bardyn et~al.}(2009)\citenamefont{Bardyn, Liew, Massar,
  McKague, and Scarani}}]{Bardyn}
\bibinfo{author}{\bibfnamefont{C.-E.} \bibnamefont{Bardyn}},
  \bibinfo{author}{\bibfnamefont{T.~C.~H.} \bibnamefont{Liew}},
  \bibinfo{author}{\bibfnamefont{S.}~\bibnamefont{Massar}},
  \bibinfo{author}{\bibfnamefont{M.}~\bibnamefont{McKague}}, \bibnamefont{and}
  \bibinfo{author}{\bibfnamefont{V.}~\bibnamefont{Scarani}},
  \bibinfo{journal}{Phys. Rev. A} \textbf{\bibinfo{volume}{80}},
  \bibinfo{pages}{062327} (\bibinfo{year}{2009}).

\bibitem[{\citenamefont{Rabelo et~al.}(2011)\citenamefont{Rabelo, Ho,
  Cavalcanti, Brunner, and Scarani}}]{Rabelo}
\bibinfo{author}{\bibfnamefont{R.}~\bibnamefont{Rabelo}},
  \bibinfo{author}{\bibfnamefont{M.}~\bibnamefont{Ho}},
  \bibinfo{author}{\bibfnamefont{D.}~\bibnamefont{Cavalcanti}},
  \bibinfo{author}{\bibfnamefont{N.}~\bibnamefont{Brunner}}, \bibnamefont{and}
  \bibinfo{author}{\bibfnamefont{V.}~\bibnamefont{Scarani}},
  \bibinfo{journal}{Phys. Rev. Lett.} \textbf{\bibinfo{volume}{107}},
  \bibinfo{pages}{050502} (\bibinfo{year}{2011}).

\bibitem[{\citenamefont{Peres}(1999)}]{Peres}
\bibinfo{author}{\bibfnamefont{A.}~\bibnamefont{Peres}},
  \bibinfo{journal}{Found. Phys.} \textbf{\bibinfo{volume}{29}},
  \bibinfo{pages}{589} (\bibinfo{year}{1999}).

\bibitem[{\citenamefont{Werner and Wolf}(2001)}]{Werner}
\bibinfo{author}{\bibfnamefont{R.~F.} \bibnamefont{Werner}} \bibnamefont{and}
  \bibinfo{author}{\bibfnamefont{M.~M.} \bibnamefont{Wolf}},
  \bibinfo{journal}{Quant. Inf. Comp.} \textbf{\bibinfo{volume}{1}},
  \bibinfo{pages}{1} (\bibinfo{year}{2001}).

\bibitem[{\citenamefont{Genovese}(2005)}]{Genovese}
\bibinfo{author}{\bibfnamefont{M.}~\bibnamefont{Genovese}},
  \bibinfo{journal}{Phys. Rep.} \textbf{\bibinfo{volume}{413}},
  \bibinfo{pages}{319} (\bibinfo{year}{2005}).

\bibitem[{\citenamefont{Horodecki et~al.}(2009)\citenamefont{Horodecki,
  Horodecki, Horodecki, and Horodecki}}]{Horodecki}
\bibinfo{author}{\bibfnamefont{R.}~\bibnamefont{Horodecki}},
  \bibinfo{author}{\bibfnamefont{P.}~\bibnamefont{Horodecki}},
  \bibinfo{author}{\bibfnamefont{M.}~\bibnamefont{Horodecki}},
  \bibnamefont{and}
  \bibinfo{author}{\bibfnamefont{K.}~\bibnamefont{Horodecki}},
  \bibinfo{journal}{Rev. Mod. Phys.} \textbf{\bibinfo{volume}{81}},
  \bibinfo{pages}{865} (\bibinfo{year}{2009}).

\bibitem[{\citenamefont{Eberhard}(1993)}]{Eberhard}
\bibinfo{author}{\bibfnamefont{P.~H.} \bibnamefont{Eberhard}},
  \bibinfo{journal}{Phys. Rev. A} \textbf{\bibinfo{volume}{47}},
  \bibinfo{pages}{R747} (\bibinfo{year}{1993}).

\bibitem[{\citenamefont{White et~al.}(1999)\citenamefont{White, James,
  Eberhard, and Kwiat}}]{White}
\bibinfo{author}{\bibfnamefont{A.~G.} \bibnamefont{White}},
  \bibinfo{author}{\bibfnamefont{D.~F.~V.} \bibnamefont{James}},
  \bibinfo{author}{\bibfnamefont{P.~H.} \bibnamefont{Eberhard}},
  \bibnamefont{and} \bibinfo{author}{\bibfnamefont{P.~G.} \bibnamefont{Kwiat}},
  \bibinfo{journal}{Phys. Rev. Lett.} \textbf{\bibinfo{volume}{83}},
  \bibinfo{pages}{3103} (\bibinfo{year}{1999}).

\bibitem[{\citenamefont{Brida et~al.}(2000)\citenamefont{Brida, Genovese,
  Novero, and Predazzi}}]{Brida}
\bibinfo{author}{\bibfnamefont{G.}~\bibnamefont{Brida}},
  \bibinfo{author}{\bibfnamefont{M.}~\bibnamefont{Genovese}},
  \bibinfo{author}{\bibfnamefont{C.}~\bibnamefont{Novero}}, \bibnamefont{and}
  \bibinfo{author}{\bibfnamefont{E.}~\bibnamefont{Predazzi}},
  \bibinfo{journal}{Phys. Lett. A} \textbf{\bibinfo{volume}{268}},
  \bibinfo{pages}{12} (\bibinfo{year}{2000}).

\bibitem[{\citenamefont{Hoeffding}(1963)}]{Hoeffding}
\bibinfo{author}{\bibfnamefont{W.}~\bibnamefont{Hoeffding}},
  \bibinfo{journal}{Journal of the American Statistical Association}
  \textbf{\bibinfo{volume}{58}}, \bibinfo{pages}{13} (\bibinfo{year}{1963}).

\bibitem[{\citenamefont{Azuma}(1967)}]{Azuma}
\bibinfo{author}{\bibfnamefont{K.}~\bibnamefont{Azuma}},
  \bibinfo{journal}{TohoKu Mathematical Journal} \textbf{\bibinfo{volume}{19}},
  \bibinfo{pages}{357} (\bibinfo{year}{1967}).

\bibitem[{\citenamefont{McDiarmid}(1989)}]{mcdiarmid:qc1989a}
\bibinfo{author}{\bibfnamefont{C.}~\bibnamefont{McDiarmid}}, in
  \emph{\bibinfo{booktitle}{Surveys in Combinatorics}}
  (\bibinfo{publisher}{Cambridge Univ. Press}, \bibinfo{year}{1989}), vol.
  \bibinfo{volume}{141} of \emph{\bibinfo{series}{London Math. Soc. Lecture
  Notes}}, pp. \bibinfo{pages}{148--188}.

\bibitem[{\citenamefont{Bahadur}(1967)}]{Bahadur}
\bibinfo{author}{\bibfnamefont{R.~R.} \bibnamefont{Bahadur}}, in
  \emph{\bibinfo{booktitle}{Proc. Fifth Berkeley Symp. on Math. Statist. and
  Prob.}} (\bibinfo{publisher}{Univ. of Calif. Press}, \bibinfo{year}{1967}),
  vol.~\bibinfo{volume}{1}, pp. \bibinfo{pages}{13--26}.

\bibitem[{\citenamefont{Ac\'in et~al.}(2005)\citenamefont{Ac\'in, Gill, and
  Gisin}}]{Acin}
\bibinfo{author}{\bibfnamefont{A.}~\bibnamefont{Ac\'in}},
  \bibinfo{author}{\bibfnamefont{R.}~\bibnamefont{Gill}}, \bibnamefont{and}
  \bibinfo{author}{\bibfnamefont{N.}~\bibnamefont{Gisin}},
  \bibinfo{journal}{Phys. Rev. Lett.} \textbf{\bibinfo{volume}{95}},
  \bibinfo{pages}{210402} (\bibinfo{year}{2005}).

\bibitem[{\citenamefont{Vardi and Lee}(1993)}]{Vardi}
\bibinfo{author}{\bibfnamefont{Y.}~\bibnamefont{Vardi}} \bibnamefont{and}
  \bibinfo{author}{\bibfnamefont{D.}~\bibnamefont{Lee}}, \bibinfo{journal}{J.
  Royal Stat. Soc. B} \textbf{\bibinfo{volume}{55}}, \bibinfo{pages}{569}
  (\bibinfo{year}{1993}).

\bibitem[{\citenamefont{Zhang et~al.}(2010)\citenamefont{Zhang, Knill, and
  Glancy}}]{Zhang}
\bibinfo{author}{\bibfnamefont{Y.}~\bibnamefont{Zhang}},
  \bibinfo{author}{\bibfnamefont{E.}~\bibnamefont{Knill}}, \bibnamefont{and}
  \bibinfo{author}{\bibfnamefont{S.}~\bibnamefont{Glancy}},
  \bibinfo{journal}{Phys. Rev. A} \textbf{\bibinfo{volume}{81}},
  \bibinfo{pages}{032117} (\bibinfo{year}{2010}).

\bibitem[{\citenamefont{Ristad}(1995)}]{Ristad}
\bibinfo{author}{\bibfnamefont{E.~S.} \bibnamefont{Ristad}}
  (\bibinfo{year}{1995}), \bibinfo{note}{\texttt{arXiv:cmp-lg/9508012}}.

\bibitem[{\citenamefont{Blume-Kohout}(2010)}]{Blume-Kohout}
\bibinfo{author}{\bibfnamefont{R.}~\bibnamefont{Blume-Kohout}},
  \bibinfo{journal}{New J. Phys.} \textbf{\bibinfo{volume}{12}},
  \bibinfo{pages}{043034} (\bibinfo{year}{2010}).

\end{thebibliography}
\end{document}